\documentclass[aip,rsi,reprint]{revtex4-1} 
\usepackage{graphicx}
\usepackage{color}
\usepackage{epstopdf}

\begin{document}
\title{A continuously pumped reservoir of ultracold atoms}

\author{J. Mahnke}
\email[]{mahnke@iqo.uni-hannover.de}
\author{I. Kruse}
\author{A. H\"uper}
\author{S. J\"ollenbeck}
\author{W. Ertmer}
\affiliation{Institut f\"ur Quantenoptik, Gottfried Wilhelm Leibniz Universit\"at Hannover, Welfengarten 1, 30167~Hannover, Germany}
\author{J. Arlt}
\affiliation{Institut for Fysik og Astronomi, Aarhus Universitet, Ny Munkegade 120, 8000 Aarhus C, Denmark}
\author{C. Klempt}
\affiliation{Institut f\"ur Quantenoptik, Gottfried Wilhelm Leibniz Universit\"at Hannover, Welfengarten 1, 30167~Hannover, Germany}

\date{\today}

\begin{abstract}
Typical sources of ultracold atoms operate with a considerable delay between the delivery of ensembles due to sequential trapping and cooling schemes. Therefore, alternative schemes for the continuous generation of ultracold atoms are highly desirable. Here, we demonstrate the continuous loading of a magnetic trap from a quasi-continuous atom beam. We achieve a steady state with ${3.8\times 10^{7}}\,$ magnetically trapped atoms and a temperature of $102\,\mu$K. The ensemble is protected from laser light sources, a requirement for its application for metrological tasks or sympathetic cooling. The continuous scheme is robust and applicable to a wide range of particles and trapping potentials. 
\end{abstract}

\pacs{}% insert suggested PACS numbers in braces on next line
%!!!
\maketitle %\maketitle must follow title, authors, abstract and \pacs
%%%%%%%%%%%%%%%%%%%%%%%%%%%%%%%%%%%%%%%%%%%%%%%%%%%%%

\section{Introduction}
Sources of cold atoms and molecules are in common use in a wide variety of research fields ranging from fundamental investigations on quantum gases~\cite{Leggett2001} to applications in precision metrology~\cite{Riehle2004}. The coldest temperatures are typically reached by combining laser cooling in a magneto-optical trap with subsequent evaporative cooling in a conservative potential~\cite{Leanhardt2003}. 

Laser cooling allows for rapid cooling of a relatively large number of atoms~\cite{Camara2014} and can be easily operated continuously~\cite{Dieckmann1998,Jollenbeck2011}. By employing sub-Doppler cooling techniques, it also allows access to temperatures as low as $1 \mu$K~\cite{Boiron1996}. Since it requires the use of near-resonant light, it is incompatible with many relevant applications such as the sympathetic cooling of molecules~\cite{Wallis2009} or continuous precision metrology. Evaporative cooling on the other hand does not require resonant light and reaches record low temperatures~\cite{Leanhardt2003}. Due to the large loss of atoms however, the final samples are relatively small and the process inherently produces ensembles sequentially. Despite large experimental efforts to refill a Bose-Einstein condensate (BEC)~\cite{Chikkatur2002}, to pump a BEC with thermal atoms~\cite{Robins2008} or to realize continuous evaporation on a beam of atoms~\cite{Lahaye2005}, no viable method for the continuous production of quantum degenerate ensembles has been found to date.

Nonetheless truly continuous sources of ultracold particles are highly desirable. In particular optical clocks - the most precise frequency standards to date - are currently limited by the Dick effect~\cite{Santarelli1998}, which could be avoided by a continuous interrogation scheme. Similar restrictions exist for inertial sensors based on atom interferometry. Moreover continuous cooling schemes would enable the sympathetic cooling of objects which cannot be cooled directly - such as molecules, atomic or molecular ions and nanoscopic solid state systems~\cite{Treutlein2012}.

Classic continuous cooling techniques such as buffer gas cooling or surface contact cooling have reached the millikelvin regime, and have enabled the generation of BECs~\cite{Fried1998}, without the need for laser cooling~\cite{Doret2009}. Moreover a wide variety of techniques are available for the generation of cold molecular beams~\cite{VanDeMeerakker2008}, including the deceleration of effusive and supersonic beams by Stark~\cite{Bethlem1999} and Zeeman~\cite{Phillips1982} slowers. More recently, it has been demonstrated that an effusive molecular beam can be decelerated significantly in a mechanical centrifuge~\cite{Chervenkov2014}. A continuous cooling scheme with minimal use of resonant laser light has been demonstrated by pumping Chromium atoms with an atom diode into a dark reservoir~\cite{Falkenau2011}. Despite this progress however, a continuous cooling scheme in the mikrokelvin regime without the application of resonant light is still not available.

A promising approach towards this goal is the continuous cooling of a slow atomic beam in a magnetic waveguide. It was shown previously that laser-cooled atoms can be efficiently coupled into such a wave guide~\cite{Lahaye2004,Olson2006} and that evaporative cooling of the beam~\cite{Lahaye2005,Reinaudi2006} is possible. Phase-space densities of $2\times 10^{-7}\,\textnormal{h}^{-3}$ have been reached in these experiments, showing the general viability of the approach. Further progress is expected by storing and accumulating these ultracold atoms in a dark reservoir of ultracold atoms.

\section{Loading mechanism for a conservative magnetic trap}
\label{chap:proposal}

\begin{figure*}[ht]
\centering
\includegraphics[width=0.9\textwidth]{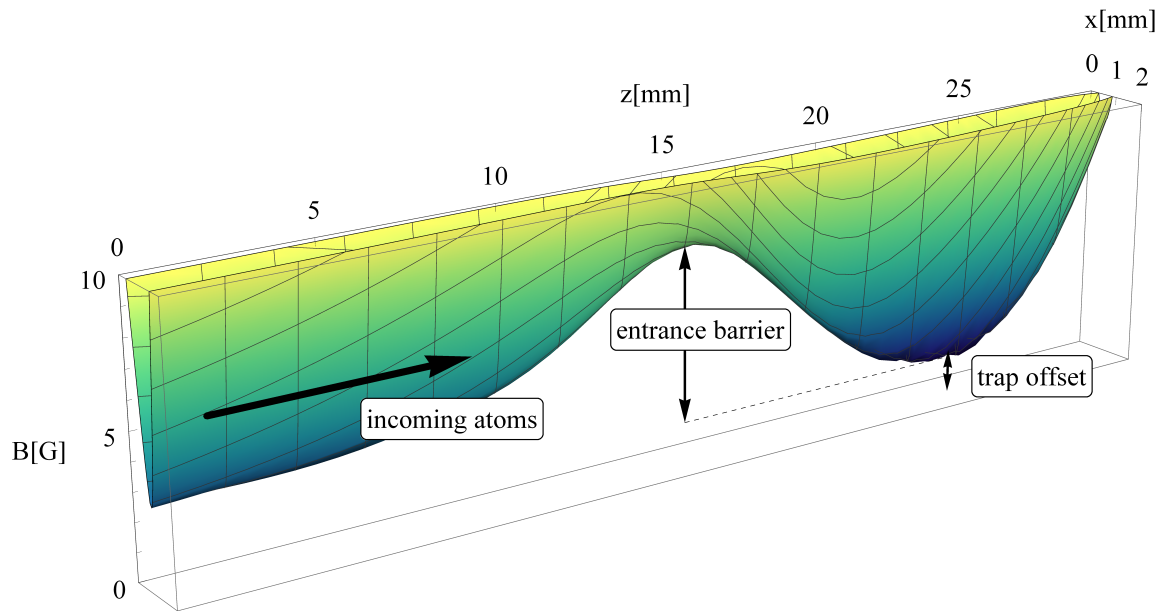}
\caption{3D plot of the magnetic trapping potential in the x-z-plane through the point of the trap minimum (see Fig. \ref{fig:block} for orientation). Incoming atoms from the left pass the entrance barrier to enter the trap where they undergo collisions. Atoms losing enough energy are trapped.}
\label{fig:3dplot}
\end{figure*}

Previous experiments on guiding and cooling of slow atom beams have led to the proposal of a magnetic trap continuously loaded by a pre-cooled atom beam~\cite{Roos2003}. In this scheme, pre-cooled atoms are guided towards the entrance barrier of an elongated magnetic trap with a finite trap depth (see Fig.~\ref{fig:3dplot}). If the atoms pass the entrance barrier, they follow the elongated magnetic potential until they are reflected by the end of the trap. The strong confinement in the radial direction ensures that most atoms collide with another atom before they reach the entrance barrier again. These collisions allow for a redistribution of the kinetic energy. Consequently, some atoms acquire a kinetic energy larger than the trap depth and escape the trap. Other atoms loose energy and stay trapped in the magnetic potential. These trapped atoms further enhance the collision probability for incoming atoms. If the trap parameters are chosen well, an equilibrium condition with a surprisingly large phase-space density may be reached.

In this paper, we present the first experimental implementation of such a continuous loading of a conservative trap. Our realization is based on a mesoscopic atom chip (see Fig.~\ref{fig:block}), a planar structure of millimeter-sized wires. The mesoscopic chip generates the magnetic fields for a three-dimensional magneto-optical trap (3D-MOT), the magnetic waveguide, and the trapping potential described above. The 3D-MOT is periodically loaded with an ensemble of atoms (section~\ref{chap:atomsource}). These ensembles are launched into the magnetic waveguide, where they overlap and produce a quasi-continuous atom beam of adjustable mean velocity (section \ref{chap:atombeamsetup}). This beam traverses an aperture which optically isolates the loading region from the trapping region where our experiments are conducted. In the trapping region, the atom beam is directed onto the elongated magnetic trap (see Fig. \ref{fig:3dplot}) with finite trap depth (section \ref{chap:trappingsetup}). Section~\ref{chap:reloadingsetup} presents our results on the continuous loading of the magnetic trap. We achieve a total number of ${3.8\times 10^{7}}$ trapped atoms at a temperature of $102\,\mu$K, corresponding to a peak phase-space density of $9 \times 10^{-8}\,\textnormal{h}^{-3}$. Such a continuously replenished ensemble of ultracold atoms without exposure to laser light presents a new tool for metrological tasks and for the sympathetic cooling of other atomic species, molecules or nanoscopic solid state systems. 

\section{Double-MOT setup as a source of laser-cooled atoms} 
\label{chap:atomsource}
The atom source used in this experiment has been described previously~\cite{Jollenbeck2011}. Briefly, it consists of a two-dimensional magneto-optical trap (2D-MOT) for $^{87}$Rb atoms, which loads a 3D-MOT situated underneath a mesoscopic atom chip.

\begin{figure*}[hbtp]
\centering
\includegraphics[width=0.9\textwidth]{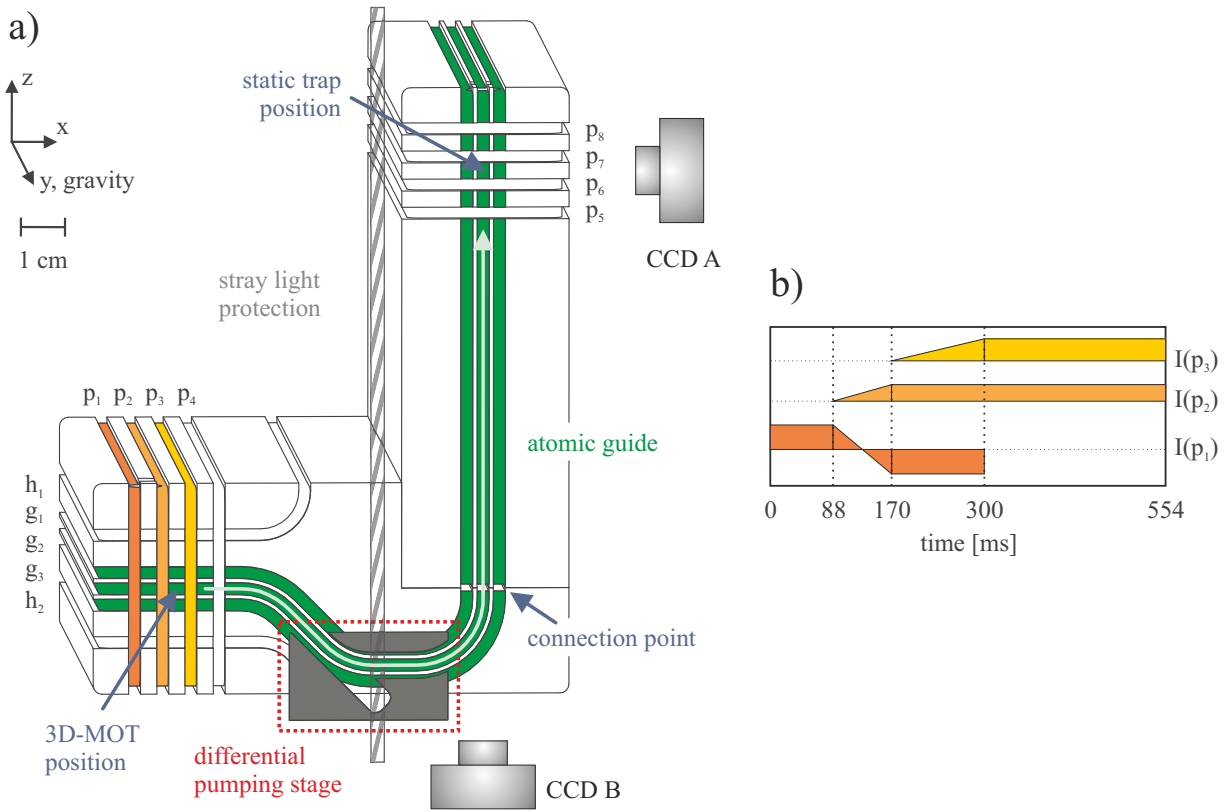}
\caption{a) Bottom view of the mesoscopic atom chip with the 3D-MOT in the lower left corner and the atomic guide (green) passing a curved channel in the partition plate (gray). The gray striped area indicates the separation of the two vacuum areas. The wires are labeled according to their intended purpose as guide wires $g_i$, perpendicular wires $p_i$ and wires for compensation of hexapole components in the MOT $h_i$~\cite{Jollenbeck2011}. The static trap is located at the end of the atomic guide and can be observed from two directions with CCD cameras. Each guide wire features an additional perpendicular lead at the connection point, that separates the wire allowing for independent control of the current through the front and rear guide wires. Therefore, the mesoscopic chip consists of two bulk copper blocks that hold and cool the wires. b) Current sequence for launching an atom cloud into the magnetic guide with ${I(p_1)=\pm 103.0}\,\textnormal{A}$, $I(p_2)={69.8}\,\textnormal{A}$ and $I(p_3)={92.6}\,\textnormal{A}$.}
\label{fig:block}
\end{figure*}

The combination of a long trapping region, high-power cooling beams and a pushing and retarding beam leads to a 2D-MOT with a flux larger than ${8.8\times 10^{10}}\,\textnormal{atoms/s}$.
All magnetic fields for trapping and manipulating atoms after the 2D-MOT are generated by the mesoscopic atom chip. Figure \ref{fig:block} shows the horizontal, planar wire structure at the bottom facet of the atom chip. The atom chip is situated outside of the vacuum system and is lowered into a L-shaped recess in the vacuum chamber. This concept allows for improvements of the atom chip without opening the vacuum system. Furthermore, the setup avoids vacuum feedthroughs for current-carrying wires and water-cooling, ensuring good vacuum conditions. The atom chip is separated from the vacuum by a ${500}\,\mu$m-thin steel foil, which is gold coated from the inside and acts as a mirror for the 3D-MOT laser beams.

The position of the 3D-MOT is marked in Fig. \ref{fig:block}. The two-dimensional quadrupole field in the y-z-plane is generated by one current-carrying wire ($g_2$) in combination with two adjacent parallel wires with opposite currents ($g_1, g_3$). Two additional wires ($h_1, h_2$) in the same plane are used to compensate distortions of the quadrupole field. Four perpendicular wires ($p_1$ to $p_4$) with currents in alternating directions generate a quadrupole field in the x-direction. A three-dimensional mirror MOT~\cite{Reichel2002} is realized by combining this quadrupole field with three laser beams, two of which are reflected on the gold-coated vacuum wall, and the third is reflected on an additional mirror inside the vacuum. The 3D-MOT is loaded from the 2D-MOT with an initial loading rate of ${8.8\times 10^{10}}\,\textnormal{atoms/s}$ and ${1.5\times 10^{10}}\,\textnormal{atoms}$ are obtained within ${300}\,\textnormal{ms}$.

After a compression and cooling phase, the atoms are optically pumped to the magnetically trappable state $F=2, m_F=2$. Afterwards,  a chip-based three-dimensional quadrupole trap can be loaded with ${1.3\times 10^{9}}\,\textnormal{atoms}$. Alternatively, the atoms can be transferred directly into a two-dimensional quadrupole guide generated by the three inner wires (marked green in Fig.\ \ref{fig:block}) with currents ${I(g_1)=105.3}\,\textnormal{A}$, ${I(g_2)=102.5}\,\textnormal{A}$ and ${I(g_3)=118.3}\,\textnormal{A}$. 

\section{Production of a quasi-continuous magnetically guided beam}
\label{chap:atombeamsetup}
A pulsed atom beam can be created by accelerating the trapped atoms in the atomic guide. The acceleration is implemented by a sequential, linear increase of the currents in the perpendicular wires $p_1$ to $p_3$ (see Fig. \ref{fig:block}b). This sequence generates a moving potential hill, accelerating the atoms towards the trapping region. In the quadrupole guide, the atoms move from the MOT region to the trapping region. These two vacuum regions are separated by a partition plate with a narrow channel for the atom beam, allowing for better vacuum conditions in the trapping region. Additionally, the partition plate shields the trapping region from the resonant light in the MOT area, enabling simultaneous operation on both sides.

After passing the channel, the atoms are guided to the trapping region. The trapping region features an absorption detection setup for the characterization of the atom beam. It includes two CCD cameras, which image the atoms from the side and along the beam.

\begin{figure}[htbp]
	\centering
	\includegraphics[width=1.\columnwidth]{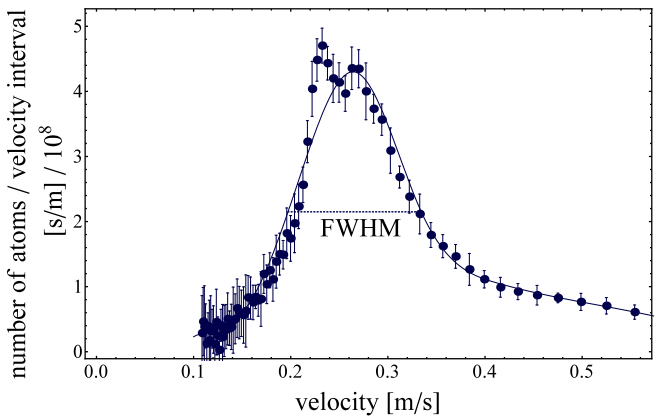}
		\caption{Atoms passing a small section of detection system A. The beam velocity, the velocity spread and the total number of atoms can be derived. The error bars represent the standard deviation of sixteen experimental realizations. The solid line is a guide to the eye.}
	\label{fig:beamarrival}
\end{figure}

The beam velocity can be measured by launching an ensemble of atoms and detecting its arrival at various positions within the $15\,$mm wide detection area of CCD camera A. A mean velocity of ${25.6}\,\textnormal{cm/s}$ is inferred.
The velocity distribution shown in Fig.~\ref{fig:beamarrival} can be extracted by taking the total traveling distance of ${20}\,\textnormal{cm}$ from the MOT into account. The nontrivial shape results from the three-step acceleration process which was automatically optimized to yield a maximal atom flux. A total number of ${8.4\times 10^{7}}\,\textnormal{atoms}$ per launch is obtained by integrating over the distribution. The width of the distribution yields a longitudinal temperature of ${93}\,\mu\textnormal{K}$. The radial temperature of ${77}\,\mu\textnormal{K}$ can be measured with a time of flight measurement in the trapping region. 

For a repetitive operation, the guide currents in the MOT area and in the trapping region can be controlled separately. As soon as the atoms cross the connection point between the two parts of the chip (see Fig. \ref{fig:block}), another cloud of atoms can be loaded in the MOT, prepared and launched. Such a generation of a quasi-continuous atom beam yields a total flux of ${7.6\times 10^{7}}\,\textnormal{atoms/s}$. 

\section{Realization of an elongated trap on a mesoscopic atom chip}
\label{chap:trappingsetup}
The mesoscopic atom chip allows for the implementation of the proposal~\cite{Roos2003} as described in section \ref{chap:proposal}. The three guide wires ($g_1$ to $g_3$) generate the atomic quadrupole guide, while the four perpendicular wires ($p_5$ to $p_8$) are used to create a three-dimensional magnetic trap with adjustable entrance barrier height. Both the trap offset and the longitudinal trap frequency can be chosen by adjusting the currents in the wires. The optimal trapping potential is obtained from an automatic multi-dimensional optimization (see section 6) and is realized with the currents ${I(p_5)=28.1}\,\textnormal{A}$, ${I(p_6)=0.5}\,\textnormal{A}$, ${I(p_7)=-3.9}\,\textnormal{A}$ and ${I(p_8)=70.9}\,\textnormal{A}$ (Fig.~\ref{fig:3dplot}).

The displayed magnetic fields are the result of a numerical integration of Biot-Savart's law~\cite{Klempt2008}. The simulation includes the influence of the supply wires. We confirmed that the wires can be approximated by infinitesimal conductors at the relevant distances. We have experimentally verified the results of the simulation by measuring the offset magnetic field at the center of the trap and the trap frequencies in the three main directions. In the experiment, the trap offset is measured by loading the trap with an evaporatively cooled ensemble in the $F=2, m_F=2$-state. Microwave spectroscopy of the transition to the $F=1, m_F=1$-state yields an offset field of ${1.15}\,\textnormal{G}$ in excellent agreement with the calculation.

The trap frequencies are measured by perturbing a cold atom cloud by small changes of the magnetic field. This leads to oscillations of the cloud in each direction. We determine the trap frequencies $\omega_{x,y,z}=2 \pi \times (171, 168, 5.6)\,$Hz by monitoring the atoms with CCD cameras A and B. The trap parameters are in good agreement with the results of the simulations $\omega_{x,y,z}=2 \pi \times (186, 185, 6.4)\,$Hz and feature the desired elongation with an aspect ratio of $30$. This elongated magnetic trap presents an ideal starting point to test the continuous loading scheme.

\section{Continuous loading of a reservoir of ultracold atoms}
\label{chap:reloadingsetup}
To realize the continuous loading mechanism, we repeatedly prepare atom clouds and accelerate them into the atomic guide, thus creating an atom beam with a flux of $7.6\times 10^{7}\,$atoms/s.  When these atoms pass the entrance barrier, the accumulation of atoms in the static magnetic trap starts.

\begin{figure*}[htbp]
	\centering
	\includegraphics[width=1.\textwidth]{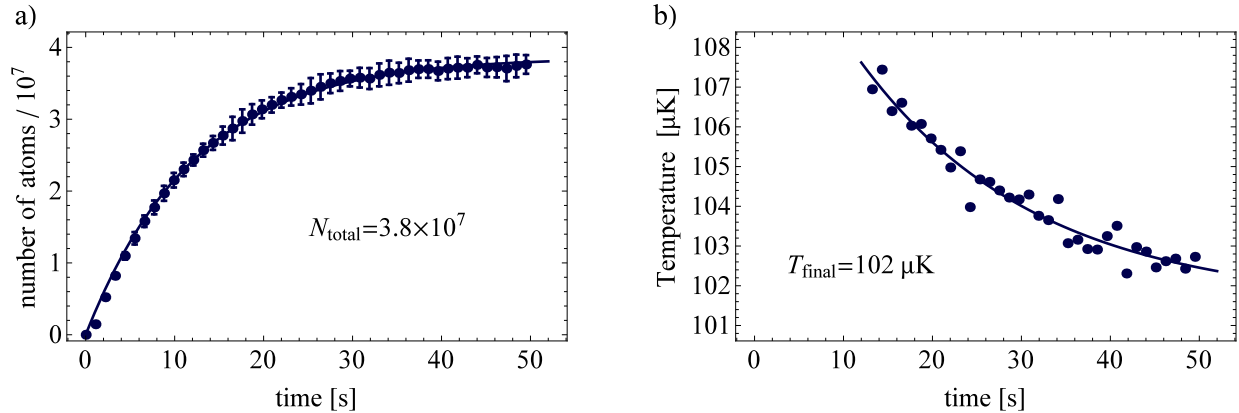}
		\caption{Quasi-continuous loading of the trap. a) number of atoms, b) temperature as a function of time. The temperature is derived from time of flight images taken at ${5}\,\textnormal{ms}$ and ${8}\,\textnormal{ms}$. An exponential decay fit to the temperature data suggests a final temperature of T$_{\textnormal{final}}=102\,\mu$K for longer loading times, while the number of atoms is fully saturated after 40\,s. The error bars represent the standard deviation of eight evaluated images.}
	\label{fig:reload}
\end{figure*}

Figure \ref{fig:reload} a) shows the number of trapped atoms as a function of loading time. The number of atoms is measured after a holding time of ${2.25}\,\textnormal{s}$ to ensure that only trapped atoms are counted. The number of atoms reaches a final value of ${3.8\times 10^{7}}\,\textnormal{atoms}$ with an initial loading rate of ${3.1\times 10^{6}}\,\textnormal{atoms/s}$. As predicted, the increase in the number of atoms is accompanied by a decrease in temperature as measured during ballistic expansion (see Fig. \ref{fig:reload} b). The temperature stabilizes at ${102}\,\textnormal{$\mu$K}$ for long loading times. We obtain a peak phase-space density of $9 \times 10^{-8}\,\textnormal{h}^{-3}$ from the recorded density profiles, the number of atoms, and the temperature. The accumulation of ultracold atoms in a conservative trap presents the key result of this publication. In the future, such a continuously pumped reservoir can be used as a steady coolant for efficient sympathetic cooling with constant cooling power.

\begin{figure}[htbp]
	\centering		\includegraphics[width=1.\columnwidth]{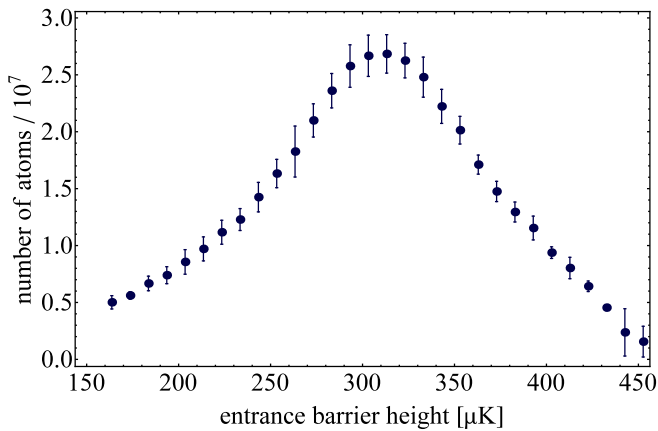}
		\caption{Number of accumulated atoms depending on the entrance barrier height after continuous trap loading for ${33}\,\textnormal{s}$. The error bars represent the standard deviation of four experimental realizations.}
	\label{fig:potentialhight}
\end{figure}

The exact configuration of the trap setup is the result of a multi-dimensional optimization of 13 parameters. Such a large number of highly correlated parameters in a noisy environment can only be  optimized efficiently by automated computer algorithms. We employ an especially well-suited algorithm called differential evolution~\cite{Geisel2013}. This algorithm was used to adjust all trap parameters, including the crucial entrance barrier height. A scan of the barrier height without further trap adjustment is shown in Fig. \ref{fig:potentialhight}. For low barrier heights, the atoms are likely to leave the trap even after a thermalizing collision. For large barrier heights, the flux of atoms that enter the trap is reduced. A balance between these two effects is found for an optimal barrier height of ${310}\,\mu\textnormal{K}$, where $65$\% of the atoms enter the trap, while the remaining low-energy atoms are reflected.

\begin{figure}[htbp]
	\centering	
	\includegraphics[width=1.\columnwidth]{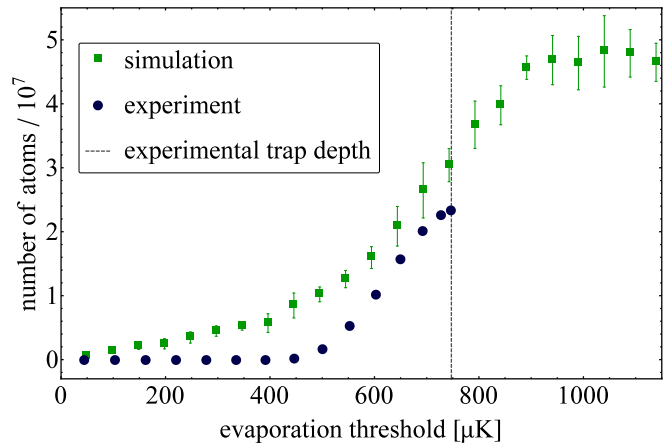}
		\caption{Accumulated atoms after 33\,s of continuous loading with a microwave evaporation threshold present during the entire trap loading time. The blue circles are experimental data and the  dashed line indicates the experimental trap depth. The green squares are data points from a molecular dynamics simulation without a trap depth limitation but otherwise equal parameters.}
	\label{fig:evaporationthreshold}
\end{figure}

An efficient accumulation of atoms requires a transverse evaporation threshold to quickly eliminate hot atoms~\cite{Roos2003}. Such a threshold is naturally realized by the finite trap depth of our magnetic chip trap of $750\,\mu$K. In addition, we have investigated the influence of the evaporation threshold by forced microwave evaporation resonant with the transition $F=2, m_F=2 \rightarrow F=1, m_F=1$. Figure \ref{fig:evaporationthreshold} shows the number of atoms obtained after $33$s of loading with continuous microwave irradiation during the entire loading process. The number of accumulated atoms increases with the evaporation threshold, until it reaches the maximally obtainable trap depth set by the trap configuration. These data suggest that the obtained results are limited by the absolute trap depth, which is technically restricted by the current limits of our mesoscopic atom chip. We employed a molecular dynamics simulation according to Ref.\cite{Roos2003} to estimate the optimal trap depth and the corresponding achievable number of atoms. The simulation shows that an increased trap depth would indeed lead to a larger number of atoms, but our realization is already quite close to the optimum. A further increase thus necessitates either a tighter radial trapping confinement or an improved incoming atom flux.

The achieved results could only be obtained due to a long lifetime of the trapped atoms. We have measured a lifetime of $240\,$s due to collisions with the background gas. This lifetime is much larger than the relevant time scale of the continuous loading process of $12\,$s. However, the continuous operation of the atom beam involves resonant laser light in the MOT area which may lead to an additional loss in the shielded trapping region. We have investigated this effect by operating the atom beam creation sequence without atoms while holding previously accumulated atoms in the static magnetic trap. Resonant laser light and external noise sources were measured to yield an additional loss of less than $20\,\%$ after a hold time of $30\,$s, and thus do not influence the performance of the continuous loading scheme. However, the switching magnetic fields in the atom beam creation sequence influence the trapped atoms more strongly: We record a nearly exponential loss with a rate of $12\,$s. Thus, the cross-talk between the two areas of the mesoscopic atom chip limits the loading time scale and, consequently, the achievable number of trapped atoms. 

\section{Conclusion}
In summary, we have demonstrated for the first time that atoms from a continuous atom beam can be accumulated in a static magnetic trap. We have achieved an equilibrium number of ${3.8\times 10^{7}}$ trapped atoms at a temperature of ${102}\,\textnormal{$\mu$K}$. The continuous loading scheme is implemented on a versatile mesoscopic atom chip. The results offer exciting perspectives for an application as a continuous source of ultracold atoms, either as a coolant for sympathetic cooling or as a resource for metrology. The presented continuous loading scheme does not exploit specific atomic properties and can thus be easily adopted to any other trappable particle. In the future, we believe that we can increase the trapping potential by a factor of two by allowing for larger currents in the atom chip. Furthermore, an improvement of the transfer efficiency into the guide will be achieved by avoiding Eddy currents during the molasses cooling, allowing for an improvement of the atom flux by a factor of $15$. For this case, the molecular dynamics simulation predicts an increase of the phase-space density by a factor of $80$. By applying a high-intensity dipole trap, the trapping confinement could be further enhanced and pushed towards the continuous operation of a quantum degenerate sample.

\section*{Acknowledgments}

We acknowledge support from the Centre for Quantum Engineering and Space-Time Research (QUEST) and from the Deutsche Forschungsgemeinschaft (Research Training Group 1729 ``Fundamentals and applications of ultra-cold matter''). J.A. thanks the Lundbeck Foundation for financial support.

\section*{References}

\bibliography{mahnke}

\end{document}